\documentstyle[11pt,epsf]{article}
\begin{document}
\textwidth 15cm
\textheight 20cm
{\noindent{\large\bf Effect of Geometrical Constraint On Conformational Properties Of A Polymer Chain}}

\vspace {.4cm}
{\noindent{\it\large Pramod Kumar Mishra}}
\vspace {.4cm}

{\noindent{\bf Department of Physics, DSB Campus, Kumaun University, Naini Tal-263 002 (Uttarakhand) India.}}
\vspace {.2cm}

{\noindent Email: pkmishrabhu@gmail.com}
\vspace {.2cm}

\noindent {\bf Abstract} :

In this paper, we analyze the effect of geometrical constraint on the conformational properties of an infinitely long linear semiflexible
polymer chain confined in-between two constraints under good solvent condition in two dimensions. 
The constraints are two impenetrable stair shaped surface and for 
two dimensional space, the surface is a one dimensional line. 
The semiflexibility of the chain is accounted by introducing a Boltzmann weight of bending energy required to produce 
each turn in the chain and good solvent condition was accounted by using self avoiding walk model of the chain. 
We have calculated exact critical value of step fugacity required for polymerization of 
an infinitely long polymer chain confined in between the constraints for different values of separation between the constraints
for directed version of the model. We have also calculated
possible maximum, minimum values of the persistent length for such chains and the maximum value of bending energy required for each turn in the chain 
for few values of separation between the constraints. 

\vspace{.2cm}
{\noindent{\bf keywords: Polymer chain, Semiflexible, Constraint, Exact Results}}

\section{Introduction}
Biomolecules ($DNA$ \& $proteins$) live in the crowded, constrained regime. Such molecules are soft object and therefore
can be easily squeezed into the spaces that are much smaller than the natural size of the molecule in the bulk. For instance,
$actin$ filaments in $eukaryotic$ cell \cite{1} or $protein$ encapsulated in $Ecoli$ \cite{2} found in nature are the
examples of confined biomolecules that serves as the basis for understanding numerous phenomenon observed in the polymer technology,
bio-technology and many other molecular processes occurring in the living cells. The conformational properties of a single
biomolecule have attracted considerable attention in the recent years due to developments in the single molecule based observations
and experiments \cite{3,4,5,6,7,8}. Under confined geometrical condition, the excluded volume effect and effect of the geometrical constraint
compete with entropy of the molecule. Therefore, geometrical constraint can modify the conformational properties 
of the molecules.

The conformational properties of a linear flexible polymer molecule under good solvent condition, confined to flat parallel walls (slit) have been studied 
for past few years; for instance, see, \cite{9,10,11,12,13,14,15} and references quoted therein. Whittington and his 
coworkers \cite{12,13,14,15} used self avoiding
walk model to study behaviour of a surface interacting flexible polymer chain confined between two parallel walls on a square lattice. They
calculated phase diagram of the polymer chain having attractive interaction with the walls by solving the
directed walk model exactly \cite{12}. Rensburg {\it et al.} \cite{15} through numerical calculation using isotropic self avoiding walk model
showed that phase diagram obtained for a surface interacting linear polymer chain confined in between two
parallel walls has qualitatively similar phase diagram to that obtained by Brak {\it et al.} \cite{12} for directed walk model of the
problem.  

Theoretical understanding of a semiflexible polymer chain confined in a cylindrical pore, rectangular shaped micro-channel, tube like narrow
channel or other kind of confined regime is also discussed in the literature to analyze  
effect of the confinement on the 
conformational behaviour of the chain; see, for instance \cite{16,17,18,19,20,21,22,23} and references therein. 
However, in present investigation, we have considered a linear 
semiflexible homopolymer chain confined in between
two one dimensional stair shaped impenetrable surface (geometrical constraint) under good solvent condition in two dimensions. 
For two dimensional space, the surface is a line and polymer chain is constrained by the such surface. 

To analyze effect of the geometrical constraint on the conformational properties of the semiflexible polymer, we have chosen
fully directed self avoiding walk model introduced by Privmann {\it et. al} \cite{24,25} and used generating function technique 
to solve the model analytically.  
The results so obtained is used to discuss behaviour of the polymer chain in constrained geometry and also to compare the 
results obtained for conformational properties of the polymer chain, when chain is in the bulk \cite{26} and there is no constraint near it.

The outline of the paper is as follows: In Sec. 2, we describe the lattice model of
fully directed self avoiding walk and used it to model a semiflexible homopolymer chain confined in between constraints (stair shaped surface). 
We have solved the model analytically to calculate exact critical value of step fugacity required for polymerization of an infinitely long linear
semiflexible polymer chain when the chain is confined in between the constraints. We have discussed the variation
of minimum critical value of step fugacity (for flexible chain) with separation between the constraints. 
We have also studied the variation of minimum and maximum
possible values of the persistent length of the flexible and stiff chains respectively when it is confined in between the constraints. 
Finally, in Sec. 3, we discuss the results obtained.

\section{Model and method}
A model of fully directed self-avoiding walk \cite{24,25} on a square lattice has been used to 
calculate conformational properties a linear semiflexible homopolymer chain confined in between two
impenetrable stair shaped surface under good solvent condition (as shown schematically in figure 1).  The directed walk model is
restrictive in the sense that the angle of bending has unique value, that is $90^{\circ}$ (for present case)
and directedness of the walk amounts to certain degree of stiffness in the 
walks of the chain because all
directions of the space are not treated equally. 
However, directed self avoiding walk model can be solved 
analytically and therefore it gives exact value of  
the conformational properties of the polymer chain.  
Since, we consider fully directed self avoiding walk ($FDSAW$) 
model and therefore, walker is allowed to take steps along $+x$, and $+y$ directions on a square lattice in between the constraints.
\begin{figure}[htbp]
\epsfxsize=10cm
\centerline{\epsfbox{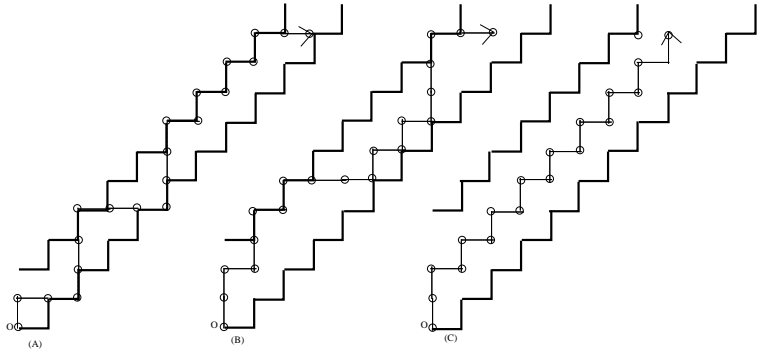}}
\caption{This figure shows a walk of an infinitely long linear semiflexible polymer chain confined in between two constraints 
(impenetrable stair shaped surface). All walks of the chain starts from a point $O$ on the constraint. 
We have shown three different cases viz. (A), (B) and (C) having separation ($n$) between the
constraints along an axis three, four and five monomers (steps) respectively. The separation between the constraints have been defined 
on the basis of the fact that how many maximum number of steps a walker can successively move along any of the $+x$ or $+y$ direction.}
\label{Figure1}
\end{figure}

The walks of the chain starts from a point $O$, located on an impenetrable surface and walker moves through out the space in between the 
two surface (as we have shown schematically in figure (1) that a walk of the polymer chain confined in between two surface for three different values
of separation ($n$) between the two surface or the constraints).  

The stiffness of the chain is accounted by associating a Boltzmann weight with bending energy for each turn in the walk of the polymer chain. 
The stiffness weight is $k${\Large(}$=exp(-\beta\epsilon_{b})$; where $\beta=\frac{1}{k_bT}$ is 
inverse of the temperature, $\epsilon_b(>0)$ is the energy associated with 
each bend in the walk of the chain, $k_b$ is Boltzmann constant and $T$ is temperature{\Large)}. 
For $k=1$ or $\epsilon_{b}=0$ the chain is said to be flexible and for 
$0<k<1$ or $0<\epsilon_{b} <\infty$ the polymer chain is said to be 
semiflexible. However, when $\epsilon_{b}\to\infty$ or $k\to0$, 
the chain has shape like a rigid rod.

The partition function of a semiflexible polymer chain can be written as, 
\begin{equation}
Z(g,k)={\sum}^{N=\infty}_{N=0}\sum_{ all\hspace{0.07cm}walks\hspace{0.07cm}of\hspace{0.05cm}N\hspace{0.05cm}steps} {g}^{N}k^{N_b}
\end{equation}
where, $N_b$ is the total number of bends in a walk of $N$ steps (monomers),
$g$ is the step fugacity of each monomer of the chain.

The partition function of an infinitely long linear semiflexible homopolymer chain confined in between the constraints 
(as shown schematically in figure 1A) can be calculated using the method of generating function
technique. The components (as shown in figure 2) of the partition function, 
$Z_3(k,g)$ (we have used here suffix three because in this case ({\it i. e.} figure 1A) maximum step that a walker can move 
successively in one particular direction is three) of the chain can be written as,
\begin{figure}[htbp]
\epsfxsize=10cm
\centerline{\epsfbox{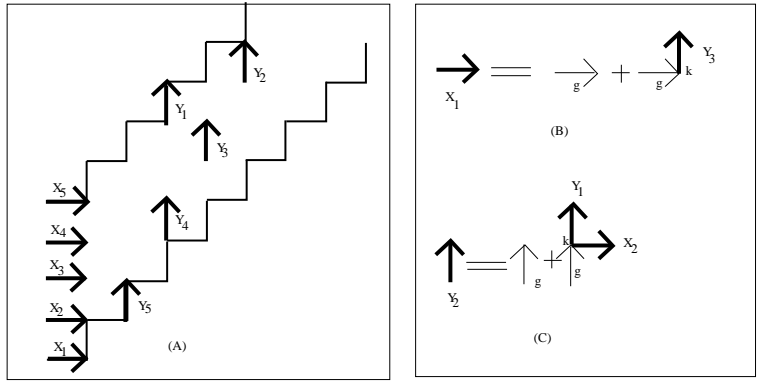}}
\caption{The components of the partition function is shown graphically in this figure. Term $X_m$ indicates sum of Boltzmann weight
of all the walks having first step along $+x$ direction, suffix $m(1\le m\le n)$ indicates maximum number of steps that walker can successively take
along $+x$ direction and $n$ is separation between the two constraints. 
Similarly, we have defined $Y_m$, where first step of the walker is along $+y$ direction. In this figure, (B) and (C)
graphically represents recursion relation for Eqs.(2\&6) respectively.}
\label{Figure2}
\end{figure}
\begin{equation}
X_1=g+kgY_3
\end{equation}
\begin{equation}
X_2=g+g(X_1+kY_2)
\end{equation}
\begin{equation}
X_3=g+g(X_2+kY_1)
\end{equation}
\begin{equation}
Y_1=g+kgX_3
\end{equation}
\begin{equation}
Y_2=g+g(kX_2+Y_1)
\end{equation}
and
\begin{equation}
Y_3=g+g(kX_1+Y_2)
\end{equation}

On solving Eqs. (2-7), we get the expression for $X_1(k,g)$ and $Y_2(k,g)$. In obtaining the expression for $X_1(k,g)$ and $Y_2(k,g)$, 
we have solved a matrix of $2nX2n$ ($n=3$, for present case {\it i. e.} figure 1A).
Thus, we have exact expression of the partition function for the semiflexible polymer chain confined between the constraints (as shown in figure 1A), 
can be written as,
\begin{equation}
Z_3(k,g)=X_1(k,g)+Y_2(k,g)=-\frac{-2g-g^2-2kg^3+2k^2g^3}{1-kg-k^2g^2-kg^3+k^3g^3}
\end{equation}
From singularity of the partition function, $Z_3(k,g)$, we obtain critical value of step fugacity required for
polymerization of an infinitely long linear semiflexible polymer chain in between the constraints (as shown in figure 1(A)). 

The persistent length of the chain can be defined using tangent-tangent correlation function 
$<{\underline t(s)}.{\underline t(s')}> \sim exp(-|s-s'|/l_p)$ \cite{27}. The tangent vector ${\underline t}(s)$ can be defined as $\partial{\underline r(s)}/\partial t$, here ${\underline r}(s)$ is parameterized in terms of arc length s of the polymer chain \cite{27}.
However, we have defined the  
persistent length $l_p$, as the average length of the polymer chain between its two successive bends \cite{26,28}.
Thus, the persistent length is calculated by us using following relation,
\begin{equation}
l_p=\frac{<N>a}{<N_b>}=(g\frac{\partial Log Z_3(k,g)}{\partial g})/(k\frac{\partial Log Z_3(k,g)}{\partial k})    \hspace{3cm} \cite{26}
\end{equation}
Where, $a$ is lattice parameter of the square lattice and we have chosen its value unity for mathematical sake.

The method discussed above can be used for different values of spacing ($n$) between the constraints and 
size of the matrix needed to solve in calculating partition of the chain
confined in between the constraints is $2nX2n$. We have calculated exact expression of the partition function for $n$ $(3\le n\le19)$.

We have plotted the maximum value of the persistent length $(l_p(Max.)=l_p(k=k_{Min.},g_c(k=k_{Min.}))$ accessible to the 
stiff polymer chain confined in between the constraints for different values of
$n$ ($3\le n\le19$) in figure (3) and found that maximum value of the persistent length scales linearly with $n$ as, $l_p=0.66272+0.40723n$. 
The value of $k_{Min.}$ is determined from singularity of the partition function $Z_3(k,g)$ and it gives least possible value of stiffness weight or 
maximum value of bending energy of the chain 
for which an infinitely long linear semiflexible chain is polymerized in between the constraints for a given value of $n$.
Since, stiffness weight
of the chain is related with bending energy $(\epsilon_b)$ as, $k=exp(-\beta\epsilon_b)$, therefore, we have calculated the maximum value of bending energy
accessible to the stiff chain confined in between the constraints from minimum value of $k(=k_{Min.})$ and plotted 
maximum value of bending energy with $n$ in figure (3). 
\begin{figure}[htbp]
\epsfxsize=10cm
\centerline{\epsfbox{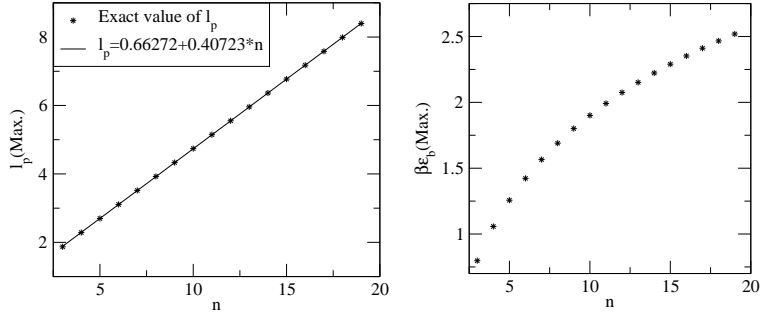}}
\caption{The exact value of the persistent length ($l_p$) is shown for different values of $n$ (the maximum number of steps that a walker can 
successively move along either $+x$ or $+y$ direction in between the constraints). We have found that maximum value of the persistent length
scales linearly with $n(3\le n\le19)$. The maximum value of the persistent length ($l_p(Max.)=l_p(k=k_{Min.},g_c(k=k_{Min.})$) 
follows a scaling relation, $l_p=0.66272+0.40723n$. 
We have also calculated the maximum value of the bending energy ($\beta\epsilon_b=-ln(k=k_{Min.})$) associated with an infinitely
long linear semiflexible chain which can be polymerized in between two constraints for different values of $n$. The maximum value of the persistent
length and bending energy corresponds to a possible stiffer chain can be polymerized in between the constraints for given values of $n$}. 
\label{Figure3}
\end{figure}

However, minimum value of the persistent length accessible to the confined chain and minimum critical value of the step fugacity accessible
to the chain is shown for various values of $n$ $(3\le n\le19)$ in figure (4).
\begin{figure}[htbp]
\epsfxsize=10cm
\centerline{\epsfbox{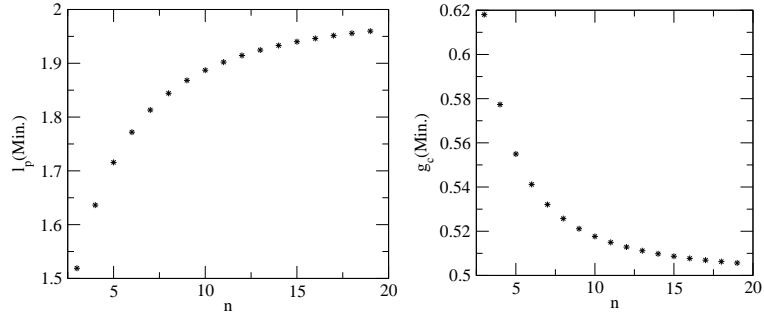}}
\caption{ In this figure, we have shown minimum value of the persistent length ($l_p(Min)=l_p(k=1,g_c(k=1)$)
and minimum critical value of step fugacity ($g_c(Min)=g_c(k=1)$)
required for polymerization of an infinitely long linear semiflexible polymer chain in between two constraints for different
values of spacings ($n$) between the constraints. In other words, we have shown critical value of step fugacity and persistent length
of a flexible polymer chain can be polymerized in between the constraints for $n(3\le n\le19)$.}
\label{Figure4}
\end{figure}
\section{Result and discussion}
We have considered an infinitely long linear semiflexible homopolymer chain confined in between two impenetrable stair shaped 
surface (constraint) in two dimensional space under good solvent condition. We have used fully directed self avoiding walk 
model to study effect of geometrical constraint imposed on the polymer chain and solved the model analytically to calculate exact 
expression of the partition function for few values of separation (n) between the constraints $3\le n\le19$. We have found that maximum
value of the persistent length scales linearly ($l_p=c+m*n$, where c=0.67745, m=0.40472 and $n\to\infty$). Thus, the persistent length of
a possible stiffer chain 
which can be polymerized in between the constraints scales linearly with spacing ($n$) between the constraints. 
We have estimated value of $c$ and $m$ for spacing ($n\to\infty$) between constraints using linear extrapolation (as shown in figure 5).
We expect that exact maximum value of the persistent length will increase linearly for all values of $n(\ge3)$. 

\begin{figure}[htbp]
\epsfxsize=10cm
\centerline{\epsfbox{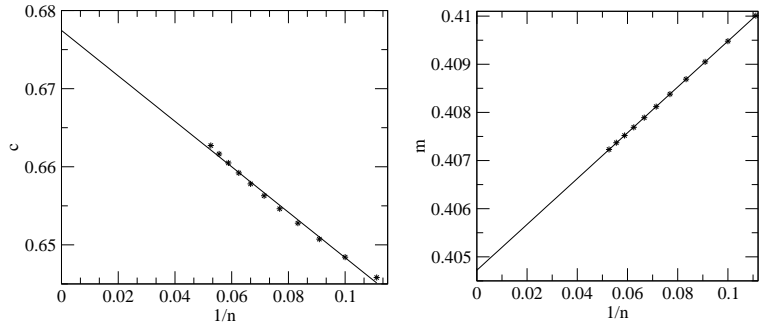}}
\caption{ We have shown in this figure different value of $c$ and $m$ for few value of $n$ and used linear extrapolation to obtain value of $c$ and $m$
for $n\to\infty$.}
\label{Figure5}
\end{figure}
We have also analyzed variation of the maximum value of the bending energy of the polymer chain confined in between the constraints and found
that it increases with
the increase of separation between the constraints for stiffer chains. The minimum value of the persistent length of a flexible polymer chain 
varies with separation between the constraints and appears to approach a value less than 2 for $n\to\infty$. 
However, critical value of step fugacity ($g_c(Min.)$=$g_c(k=1)$) required for polymerization of an infinitely long linear 
flexible polymer chain in between the constraints approaches to 0.5 as $n\to\infty$. The minimum critical value of step fugacity is 
0.5 ({\it i. e.} for $k=1$) for a long linear polymer chain when 
chain is modeled as a fully directed self avoiding walk on a two dimensional square lattice 
in absence of the constraints, {\it i. e.} $g_c(k)=\frac{1}{1+k}$ \cite{26}. 
Since, $g_c(Min.)$ is critical value of step fugacity which is required for polymerization of an infinitely long linear flexible polymer chain
on a two dimensional square lattice for fully directed self avoiding walk model and whence remain un-affected due to geometrical 
constrain imposed by stair shaped surface as $n\to\infty$. 

However, minimum value of the persistent length is less than 2 for $n\to\infty$ while the persistent length ($l_p=1+k^{-1}$) \cite{26} has value 2
for a flexible polymer chain ({\it i. e.} for $k=1$) when there is no constraints near the chain. 
The persistent length is a local property, therefore, its value depends on the type of lattice 
(square, rectangular etc.) chosen to model the chain, space dimensionality and on the fact that polymer chain is ideal or self avoiding type. 
We have found that the value of $l_p(Min.)$ is less than 2 for $n\to\infty$ and this value is different from 2, when there is no constraints near the chain
because for $n\to\infty$ there is
still one constraint near the chain and confining to it or not allowing it to occupy space below the constraint at which the chain is grafted. 


\small
\begin{thebibliography}{99}
\bibitem{1} S. Koster, D. Steinhauser and T. Pfohl, {\it Brownian motion of actin filaments in confining micro-channels}, 
J. Phys.: Cond. Matt. 17 (2005), pp. S4091.
\bibitem{2} G. Morrison and D. Thirumalai, {\it The shape of a flexible polymer in a cylindrical pore}, 
J. Chem. Phys. 122 (2005), pp. 194907-194911.
\bibitem{3} {\it Structure and dynamics of confined polymers} (ed. J. J. Kasianowicz {\it et. al}, Kluwer Academic Publishesr, Dordrecht) (2002).
\bibitem{4} S. E. Henrichson, M. Misakian, B. Bobertson and J. J. Kasianowicz, {\it Driven $DNA$ transport into an asymmetric
nanometer-scale pore}, Phys. Rev. Lett. 85 (2000), pp. 3057-3060.
\bibitem{5} D. Nykypanchuk, H. H. Strey and D. A. Hoagland, {\it Singe molecule visualizations of polymer partitioning
within model pore geometries}, Macromolecules 38 (2005), pp. 145-150.
\bibitem{6} M. Ichikawa, Y. Matsuzawa and K. Yoshikawa, {\it Entrapping polymer chain in light well under good solvent
condition}, J. Phys. Soc. J. 74 (2005), pp. 1958-1961.
\bibitem{7} C. H. Reccius, J. T. Mannion, J. D. Cross and H. G. Craig-head, {\it Compression and free expanssion of a single $DNA$
molecules in nanochannels}, Phys. Rev. Lett. 95 (2005), pp. 268101-268104. 
\bibitem{8} W. Reisner, K. J. Morton, R. Riehn, Y. M. Wang, Z. Yu. M. Rosen, J. C. Sturm, S. Y. Chou, E. Frey and R. H. Austin, 
{\it Statics and dynamics of a single $DNA$ molecules confined in nanochannels}, Phys. Rev. Lett. 94 (2005), pp. 196101-196104.
\bibitem{9} E. A. DiMarzio and R. J. Rubin, {\it Adsorption of a chain between two plates}, J. Chem. Phys. 55 (1971) pp. 4318-4336. 
\bibitem{10} P. G. de. Gennes, {\it Scaling concepts in polymer physics}, (Cornell University Press, Ithaka) (1979).
\bibitem{11} E. Eisenriegler, {\it Polymer near surfaces}, (World Scientific, Singapore) (1993).
\bibitem{12} R. Brak, A. L. Owczarek, A. Rechnitzer and S. G. Whittington, {\it A directed walk model of a long chain polymer
in a slit with attractive walls}, J. Phys. A 38 (2005), pp. 4309-4325. 
\bibitem{13} E. J. Janse van Rensburg, E. Orlandini, and S. G. Whittington, {\it Self avoiding walks in a slab: rigorous results}, 
J. Phys. A 39 (2006), pp. 13869-13903.
\bibitem{14} R. Brak, G. K. Iliev, A. Rechnitzer and S. G. Whittington, {\it Mtzkin path models of long chain polymers in slits}, J. Phys. A: Math. Theor. 
40 (2007), pp. 4415-4437. 
\bibitem{15} E. J. Janse van Rensburg, E. Orlandini, A. L. Owczarek, A. Rechnitzer and S. G. Whittington, 
{\it Self-avoiding walks in a slab with attractive walls}, J. Phys. A 38 (2005), pp. L823-L828. 
\bibitem{16} L. Harnau and P. Reineker, {\it Equilibrium and dynamical properties of semifexible chain molecules with confined
transverse fluctuations}, Phys. Rev. E 60 (1999), pp. 4671-4676. 
\bibitem{17} D. J. Bicout and T. W. Burkhardt, {\it Simulation of a semiflexibe polymer in a narrow cylindrical pore}, J. Phys. A: Math. Gen. 
34 (2001), pp. 5745.
\bibitem{18} Y. Yang, T. W. Burkhardt, and G. Gompper, {\it Free energy and extension of a semiflexible polymer in cylindrical confining geometries}, 
Phys. Rev. E 76 (2007), pp. 011804-011810. 
\bibitem{19} T. W. Burkhardt, Y. Yang, and G. Gompper,{\it Fluctuations of a long, semiflexible polymer in a narrow channel}, 
arXiv:[cond-mat.soft]:1008.1594 (2010).
\bibitem{20} P. Levi and K. Mecke, {\it Radial distribution function for semiflexible polymers confined in microchannels}, 
Europhys. Lett. 78 (2007), pp. 38001. 
\bibitem{21} F. Wagner, G. Lattanzi, and E. Frey, {\it Conformations of confined biopolymers}, Phys. Rev. E 75 (2007), pp. 050902-050905.
\bibitem{22} P. Cifra, Z. Benkova, and T. Bleha, {\it Chain extension of $DNA$ confined in channels}, J. Phys. Chem. B 113 (2009), pp. 1843-1851. 
\bibitem{23} P. Cifra, {\it Channel confinement of flexible and semiflexible macromolecules}, J. Chem. Phys. 131 (2009), pp. 224903-224909;
P. Cifra and T. Bleha, {\it Shape transition of semiflexible macromolecules confined in channel and cavity}, Eur. Phys. J. E 32 (2010), pp. 873-279.
\bibitem{24} V. Privman and H. L. Frisch, {\it Nonuniversality of the single chain rod to coil transition scaling functions: 
Exact results for directed walks}, J. Chem. Phys. 88(1) (1988), pp. 469-474. 
\bibitem{25} Privman V. and Svrakic N. M., 
{\it Directed Models of Polymers, Interfaces, and Clusters: Scaling and Finite-Size Properties} (Springer, Berlin) (1989). 
\bibitem{26} P. K. Mishra, S. Kumar and Y. Singh, {\it A simple and exactly solvable model for a semiflexible polymer chain 
interacting with a surface}, Physica A 323 (2003), pp. 453-465.
\bibitem{27} O. Kratky, G. Porod, {\it Rontgenuntersuchung geloster Fadenmolekule}, Recl. Trav. Chim. Pays-Bas 68 (1949), pp. 1106-1123; H. Yamakawa, {\it Modern Theory of Polymer Solutions}, (Harper and Row, New York) (1971).
\bibitem{28} P. K. Mishra, {\it The divergence of persistent length of a semiflexible homopolymer chain in the stiff chain limit}, (to be published).
\markboth{Effect Of Geometrical Constraint On Conformational Properties Of A Polymer Chain; P. K. Mishra}{Phase Transitions}
\end{thebibliography}
\end{document}